\documentclass[conference]{IEEEtran}

\usepackage{amsmath}
\usepackage{bbm}
\usepackage{amssymb}

\usepackage{braket}
\usepackage{graphicx}
\usepackage{cite}
\usepackage{epstopdf}
\usepackage{mathtools}
\usepackage{url}
\usepackage{bbold}
\usepackage{xcolor}

\usepackage[top=0.7in, left=0.68in, bottom=1.05in, right=0.68in]{geometry}

\newcommand{\tr}{\text{Tr}}

\newcommand{\bigchi}{\makebox{\large\ensuremath{\chi}}}

\title{Enhancing Continuous Variable Quantum Teleportation using Non-Gaussian Resources}

\author{
	 \IEEEauthorblockN{E. Villase\~nor and R. Malaney}\\
	 \IEEEauthorblockA{School of Electrical Engineering  \& Telecommunications,\\
		 The University of New South Wales, Sydney, NSW 2052, Australia.\\
     }
}

\begin{document}

\maketitle

\IEEEpeerreviewmaketitle

\begin{abstract}
Continuous Variable (CV) non-Gaussian resources are fundamental in the realization of quantum error correction for CV-based quantum communications and CV-based computing. In this work, we investigate the use of CV non-Gaussian states as quantum teleportation resource states in the context of the transmission of coherent and squeezed states through noisy channels.
We consider an array of different non-Gaussian resource states, and compute the fidelity of state teleportation achieved for each resource. Our results show that the use of non-Gaussian states presents a significant advantage compared to the traditional resource adopted for CV teleportation; the Gaussian two-mode squeezed vacuum state. In fiber-based quantum communications, the range of quantum teleportation is increased by approximately $40\%$ via the use of certain non-Gaussian states. In satellite-to-ground quantum communications, for aperture configurations consistent with the Micius satellite, the viable range of quantum teleportation is increased from 700 km to over 1200 km. These results represent a significant increase in the performance of pragmatic and realizable quantum communications in both terrestrial and space-based networks.
\end{abstract}

\section{Introduction}
The use of continuous variable (CV) quantum information is a prominent approach for quantum information protocols \cite{GaussianQuantumInformation}. Applications based on the transmission of CV quantum signals such as quantum key distribution (QKD) are particularly promising \cite{.5kmCVQKD}. One reason for this is the `off-the-shelf' availability of the devices required to send, receive, and measure CV quantum signals.  However, achieving practical and global quantum communications involves difficult technological challenges.

Currently, there exist two main approaches for constructing a global quantum communications network. One is a terrestrial implementation via optical fiber, and the second is through satellites in space.
An optical fiber implementation is able to use a similar infrastructure to the one used by the classical internet. However, fiber implementations suffer an exponential increase of loss with respect to the distance, making long-distance transmission challenging \cite{repeaters}.
On the other hand, in satellite quantum communications the main source of noise (via loss) arises in the atmospheric quantum channels connecting the satellites to Earth. If the loss due to the atmospheric channels can be dealt with, then satellites in space represent a practical approach to achieve global quantum communications.
Nevertheless, reliable transfer of quantum signals by means of satellites requires highly advanced technology and infrastructure.
While satellites have been used in quantum communication protocols using single photons \cite{liao2017satellite, yin2017satellite}, there have not been any satellite experimental realizations yet using CV quantum information. However, there is a growing interest in this latter approach, with several recent studies focusing on the satellite-to-ground transmission of quantum signals to perform CV-QKD \cite{hosseinidehaj2018satellite, Pirandola:20, FeasibilityDownlinkCVQKD}.

A key element in long-distance quantum communications is the transmission and distribution of entanglement resources, which can later be used to transmit quantum states via quantum teleportation \cite{PhysRevA.49.1473, braustein1998teleportationCV}. Recently, CV quantum teleportation has been demonstrated experimentally over 6 km \cite{Huoeaas9401}.
For quantum teleportation to be successful it is essential to maximize the amount of entanglement in the resource state to increase the quality (fidelity) of the teleported state.
Remarkably, states commonly referred to as non-Gaussian states have shown higher amounts of entanglement (at a given energy) in comparison to their Gaussian counterparts \cite{CVdistillation, scissors}.
In CV-QKD, non-Gaussian states have been successfully used directly to increase key rates
\cite{CVdistillation, Hemultimode}, including in the context of space-based CV-QKD \cite{MingjianPAPS, ScissorsQKD}.
Non-Gaussian states have also been shown to allow for long-distance state transmission by means of quantum repeaters \cite{repeaters, PhysRevResearch.2.013310}.

In this work, we consider the use of two-mode non-Gaussian states as entangled resource states in quantum teleportation over noisy channels (channels with loss and excess noise) to enhance the transmission of quantum states. We focus on several types of non-Gaussian resource states generated by means of the operations: {\it photon subtraction} (PS), {\it photon addition} (PA), {\it photon catalysis} (PC), and {\it quantum scissors} (QS). Moreover, we also consider the more general {\it squeezed Bell-like} (SB) states introduced in \cite{nonGaussianTeleportation}, as a resource state.

Previous studies have analyzed the fidelity of teleported
coherent states using non-Gaussian resource states \cite{nonGaussianTeleportation, nonGaussianTeleportation2, DellAnno1, PhysRevA.91.063832, PhysRevA.82.062329, villasenor2021enhanced}. In \cite{nonGaussianTeleportation}, the
SB state is introduced, showing improved fidelities relative to the Gaussian two-mode squeezed vacuum (TMSV) state.
This increase in fidelity is shown to still hold when
the resource states are subject to the effects of noisy channels
\cite{nonGaussianTeleportation2, DellAnno1}. In \cite{PhysRevA.91.063832}, it is shown that non-Gaussian resource states created from the symmetrical application of PS lead to an increase in
the fidelity of teleported coherent states. Moreover, the
sequential use of non-Gaussian operations PA and PS has also
been shown to create resource states that enhance quantum
teleportation, especially when such resource states are subject
to high loss \cite{villasenor2021enhanced}. However, the work of \cite{villasenor2021enhanced} did not optimize the teleportation fidelity over all the available parameter space. It is the main contribution of this current work to remedy this issue. That is, we carry out a full optimization analysis of sequential PA and PS operations to create enhanced teleportation outcomes for coherent states. We then compare our new optimized outcomes with those obtained from the use of PS, PA, PC, QS, and SB resource states. Beyond previous works, our analysis also includes the use of squeezed states as input states in teleportation. Our main conclusion is that optimized sequential PA and PS operation delivers substantial improvement in the distance over which CV quantum teleportation is feasible; for both fiber and satellite-based real-world implementations.

\section{Continuous variable quantum teleportation}
\begin{figure}
\centering
\includegraphics[width=.46\textwidth]{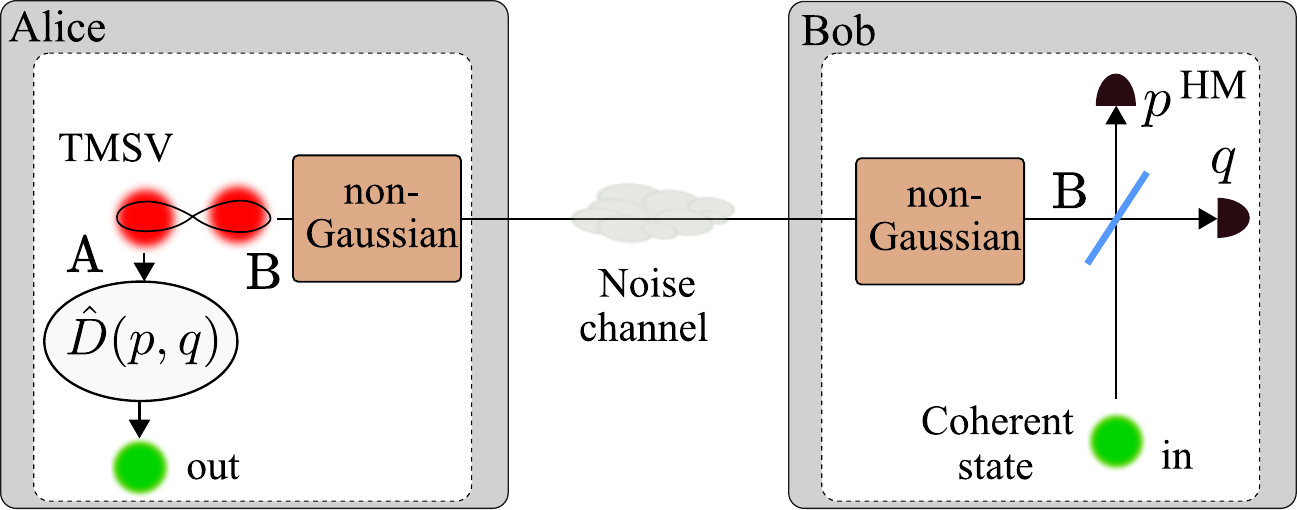}
\caption{CV quantum teleportation via a noisy channel. A non-Gaussian
operation can be applied either transmitter-side or receiver-side to create a non-Gaussian resource state. This state then forms the teleportation channel.}
\label{fig:teleportation}
\end{figure}

Quantum teleportation using CV quantum information was first introduced in \cite{PhysRevA.49.1473, braustein1998teleportationCV}. As shown in Fig \ref{fig:teleportation}, quantum teleportation is done between two parties, commonly referred to as Alice and Bob, using an entangled bipartite resource state, $\hat{\rho}_\mathrm{AB}$. Part $\mathrm{B}$ of the entangled state is transmitted by Alice to Bob via a quantum channel. After receiving part $\mathrm{B}$, Bob will use a beam-splitter (BS) to mix $\mathrm{B}$ with the state to the be teleported, $\ket{\psi}_\mathrm{in}$,
and perform homodyne measurements on both outcomes of the BS. The measurement results are then broadcast to Alice who uses these results to apply a displacement operation, $\hat{D}(q, p)$,
on part $\mathrm{A}$ of the entangled state, obtaining the teleported state, $\ket{\psi}_\mathrm{out}$. Conveniently, using the Wigner characteristic function (CF) formalism the resulting state of the teleportation protocol can be computed as the product of the CFs of the input state and the entangled resource, as shown in \cite{MarianMarian}. This result was later expanded to include the effects of non-unit efficiency homodyne measurements \cite{DellAnno1}, such that the CF of $\ket{\psi}_\mathrm{out}$ is given by
\begin{align}
  \bigchi_{\mathrm{out}}(\xi) =& \bigchi_{\mathrm{in}}(g \eta \xi) \nonumber \\
  &\times \bigchi_\mathrm{AB}(\xi, g \eta  \xi^*) e^{-\frac{|\xi|^2}{2} g^2 (1 - \eta^2) },
  \label{eq:marian}
\end{align}
where $\xi$ is a complex argument, $g$ is the gain parameter, and $\eta^2$ the efficiency of the homodyne measurements.

The CF of any $n$-mode quantum state $\hat{\rho}$ is defined as
\begin{align}
\bigchi(\xi_1, \xi_2, ..., \xi_n) = \tr\{\hat{\rho} \hat{D}(\xi_1)\hat{D}(\xi_2)...\hat{D}(\xi_n)\},
\end{align}
where $\hat{D}$ is the displacement operator,
\begin{align}
\hat{D}(\xi_i) = e^{\xi_i \hat{a}^{\dag}_i - \xi_i^* \hat{a}_i},
\label{eq:displacement}
\end{align}
with $\hat{a}_i$ and $\hat{a}^{\dag}_i$ are the annihilation and creation operators of mode $i$, respectively.

As is the case in any non-ideal quantum channel, the transmission of $\mathrm{B}$ via the channel unavoidably introduces noise to the resource state, thus, effectively decreasing the effectiveness of the teleportation. This means that higher amounts of noise result in larger differences between the outcome state and the input state, as quantified by the fidelity.
The effects of a noisy channel are described by means of two parameters; the transmissivity, $T$, and the excess noise, $\epsilon$. For the bipartite state, $\hat{\rho}_\mathrm{AB}$, where only mode $\mathrm{B}$ is transmitted through the noisy channel, the corresponding CF is,
\begin{align}
\bigchi^\mathrm{noise}_\mathrm{AB}(\xi_{\mathrm{A}}, \xi_{\mathrm{B}})&=\exp{\Big[-\frac{1}{2}(\epsilon + 1 - T)|\xi_{\mathrm{B}}|^2\Big]} \nonumber  \\
& \times ~\bigchi_\mathrm{AB}(\xi_{\mathrm{A}}, \sqrt{T} \xi_{\mathrm{B}}).
\label{eq:CF_loss}
\end{align}

Using Eq. \ref{eq:marian} and  Eq. \ref{eq:CF_loss} we can obtain the CF of the output state of the teleportation via a noisy channel. Now, in order to quantify the effectiveness of teleportation we compute the fidelity between the output and input states. The fidelity represents a measure of the closeness between the two states, in the CF formalism it is computed as,
\begin{align}
\mathcal{F} = \frac{1}{\pi} \int d^2 \xi \bigchi_\mathrm{in}(\xi) \bigchi_\mathrm{out}(-\xi).
\label{eq:fidelity}
\end{align}

In this work, two types of input states are considered in the teleportation: coherent states and squeezed states. Coherent states are defined as $\ket{\alpha} = \hat{D}(\alpha)\ket{0}$, with $\alpha \in \mathbb{C}$. Squeezed states are defined as
\begin{align}
\ket{s} =
\hat{S}(s)\ket{0}= e^{\frac{1}{2}(s^* \hat{a}^2 - s\hat{a}^\dag{}^2 )} \ket{0},
\end{align}
where $s{=}\varsigma e^{i \varphi}$ is the squeezing of the state.
Ultimately, the fidelity between output and input states will be highly dependent on the value of $\alpha$ or $s$. To address this issue, we will use instead the mean fidelity over an ensemble of states. The ensemble is specified by following the distribution,
\begin{align}\label{eq:probcoherent}
P(x) = \frac{1}{\sigma \pi} \exp\Big(-\frac{|x|^2}{\sigma}\Big),
\end{align}
with $\sigma$ the variance of the distribution, and ${x}{=}\{\alpha, s\}$.
This allows us to define the average fidelity over the ensemble as,
\begin{align}\label{eq:avefidelity}
\bar{\mathcal{F}}  = \int d{x}^2 P(x) \mathcal{F}(x),
\end{align}
where $\mathcal{F}(x)$ represents the fidelity when a specific $\ket{x}$ is used as the input state.

\subsection{Entangled resources}
The entangled resource states to be used in constructing the teleportation channel are now introduced. First,
consider a TMSV state. This can be defined through the application of the two mode squeezing operator, $\hat{S}_{12}(\varrho)$, to the vacuum state, viz.,
\begin{align}
  \ket{\psi}_\text{TMSV} = \hat{S}_{12}(\varrho) \ket{0,0} = e^{\varrho^* \hat{a}_1 \hat{a}_2 - \varrho \hat{a}^\dagger_1 \hat{a}^\dagger_2}\ket{0,0} ,
\end{align}
where $\varrho{=}re^{i \phi}$ is the squeezing of the TMSV state.
The CF of a TMSV state can be conveniently written using the following Bogoliuvov transformation:
\begin{align}
&\hat{S}_{12}(\varrho) \hat{D}(\xi_1) \hat{D}(\xi_2) \hat{S}^\dag_{12}(\varrho) = \hat{D}(\xi_1') \hat{D}(\xi_2'), \nonumber \\
&\xi_i' = \cosh(r)\xi_i + e^{i\phi} \sinh(r) \xi_j^* ~~~~ i,j = 1,2; ~~ i\neq j.
\label{eq:Bogo}
\end{align}
Using this transformation the CF of a TMSV state  can be written as,
\begin{align}
\bigchi_\mathrm{TMSV}(\xi_{\mathrm{A}}, \xi_{\mathrm{B}}) = \exp \Big[ -\frac{1}{2}\Big(|\xi_\mathrm{A}'|^2 + |\xi_\mathrm{B}'|^2 \Big) \Big].
\end{align}
Here (and throughout this work), we assume $\hbar{=}2$.

\begin{figure}
\centering
\includegraphics[width=.46\textwidth]{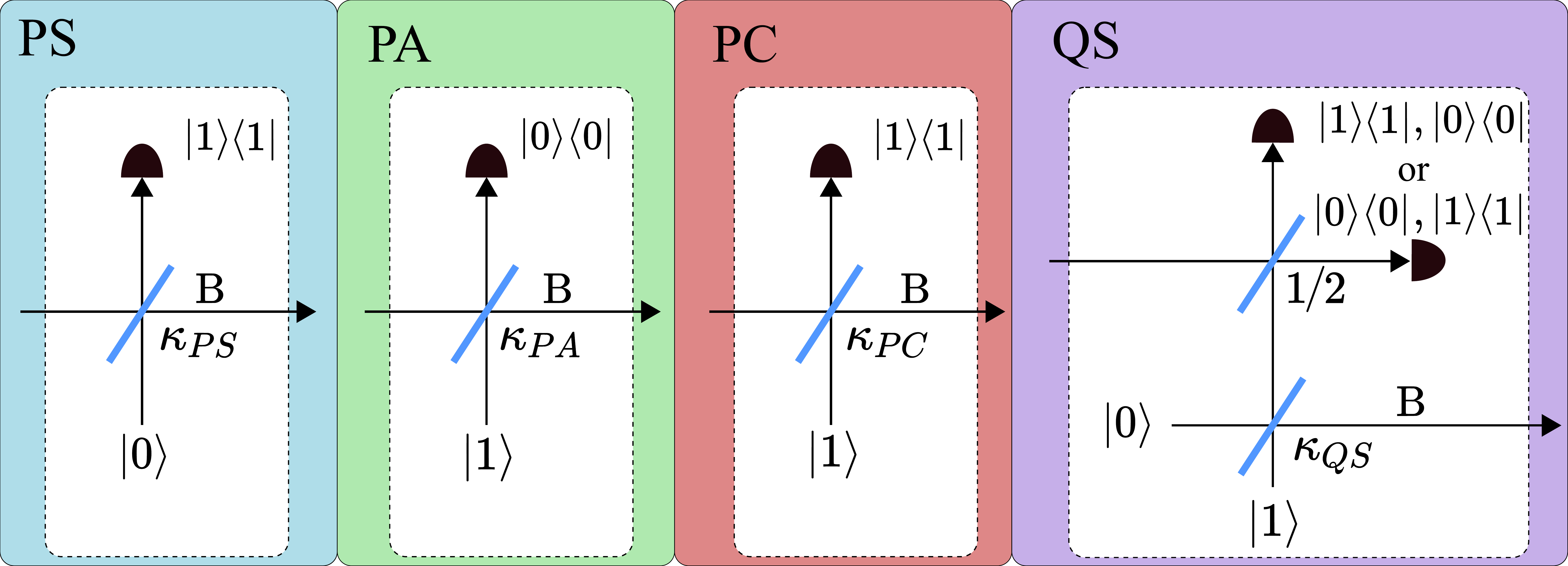}
\caption{Non-Gaussian operations used to generate a non-Gaussian resource state to be used in the teleportation channel. The operations are applied on a single mode of a TMSV state.}
\label{fig:operations}
\end{figure}

With a TMSV state as the starting point, we can obtain certain non-Gaussian states by using non-Gaussian operations. Non-Gaussian operations can be constructed with beam-splitters and by the addition or absorption of single or multiple photons. We focus on the operations, shown in Fig. \ref{fig:operations}, the operations PA-PS and PS-PA, created by the successive application of PS and PA, and the SB states.

In this work, we consider that only a single photon is added/subtracted/replaced in the respective non-Gaussian operations.
Additionally, non-Gaussian operations can be applied either transmitter-side or receiver-side, that is, before or after the effects of the noisy channel. We have investigated both cases for each operation and have found that the highest fidelities for the PS, PA, and PC operations are obtained when the operation is applied receiver-side. For the QS operation, however, the highest fidelities are obtained when the operation is applied to mode B (the mode to be sent through the channel) at the transmitter-side. These results are consistent with previous studies \cite{MingjianPAPS, 9024548}. Henceforth, we will only focus on the specific cases (transmitter-side or receiver-side) where the operations yield the highest fidelities. The transmissivity of the beam-splitter involved in each respective operation is determined by $\kappa_o$ with $o{\in}\{PS, PA, PC, QS\}$. Finally, we assume a quantum memory is available at the receiver and transmitter, such that the non-Gaussian resource states forming the teleportation channel can be prepared and stored to be used on demand.\footnote{In the absence of a quantum memory there exists a trade-off. Non-Gaussian operations have non-unity success probability; a point that needs to be taken into account in any protocol.}

\subsubsection{Photon addition, photon subtraction, and photon catalysis}
In the following we apply the procedures of \cite{villasenor2021enhanced} for determining CFs.
The {\it unnormalized} CF of the state resulting from the application of the PS operator, $\hat{O}_\mathrm{PS}$, to mode $\mathrm{B}$ of the TMSV state after it has been transmitted through the noisy channel is,
\begin{equation}
	\begin{aligned}
		\bigchi'_{\mathrm{PS}}(&\xi_{\mathrm{A}}, \xi_{\mathrm{B}})
		=\operatorname{Tr}\left\{\hat{O}_{\mathrm{PS}}\hat{\rho}_\mathrm{TMSV}^\mathrm{noise} \hat{O}_{\mathrm{PS}}^\dagger \hat{D}(\xi_{\mathrm{A}})\hat{D}(\xi_{\mathrm{B}})\right\}\\
		=&\frac{\kappa_\mathrm{PS}-1}{\kappa_\mathrm{PS}}
		\exp{\Big(-\frac{|\xi_{\mathrm{B}}|^2}{2}\Big)}\\
		&\times
		\frac{\partial^2}{\partial \xi_{\mathrm{B}} \partial \xi_{\mathrm{B}}^*}
		\bigg[
		\exp{\Big(\frac{|\xi_{\mathrm{B}}|^2}{2}\Big)}
		f(\xi_{\mathrm{A}}, \xi_{\mathrm{B}}, \sqrt{\kappa_\mathrm{PS}})
		\bigg],
	\end{aligned}
\end{equation}
where
\begin{equation}
	\begin{aligned}
		&f(\xi_{\mathrm{A}}, \xi_{\mathrm{B}}, \sqrt{\kappa_\mathrm{PS}})=\int \frac{d \xi^2}{\pi (1-\kappa_\mathrm{PS})} \bigchi_\mathrm{TMSV}^\mathrm{noise}(\xi_{\mathrm{A}}, \xi)\\
		&\times
		\exp{
			\Big[
			\frac{1+\kappa_\mathrm{PS}}{2(\kappa_\mathrm{PS}-1)}(|\xi|^2+|\xi_{\mathrm{B}}|^2)} +
			\frac{\sqrt{\kappa_\mathrm{PS}}}{\kappa_\mathrm{PS}-1}
			(\xi_{\mathrm{B}}\xi^*+\xi_{\mathrm{B}}^*\xi)
			\Big].
	\end{aligned}
\end{equation}

In a similar manner, the unnormalized CF after the PA operator is applied is,
\begin{equation}
	\begin{aligned}
	\bigchi'_{\mathrm{PA}}(\xi_{\mathrm{A}}, & \xi_{\mathrm{B}})
		=(\kappa_\mathrm{PA}-1)
		\exp{\Big(\frac{|\xi_{\mathrm{B}}|^2}{2}\Big)}\\
		&\times
		\frac{\partial^2}{\partial \xi_{\mathrm{B}} \partial \xi_{\mathrm{B}}^*}
		\bigg[
		\exp{\Big(-\frac{|\xi_{\mathrm{B}}|^2}{2}\Big)}
		f(\xi_{\mathrm{A}}, \xi_{\mathrm{B}}, \sqrt{\kappa_\mathrm{PA}})
		\bigg].
	\end{aligned}
\end{equation}
When the PC operator is applied, the unnormalized CF is,
\begin{equation}
	\begin{aligned}
		\bigchi'_{\mathrm{PC}}(\xi_{\mathrm{A}}, & \xi_{\mathrm{B}})
		=q^2
		\exp{\Big(\frac{|\xi_{\mathrm{B}}|^2}{2}\Big)}
		\frac{\partial^2}{\partial \xi_{\mathrm{B}} \partial \xi_{\mathrm{B}}^*}
		\bigg\lbrace
		\exp{\Big(-|\xi_{\mathrm{B}}|^2\Big)}\\
		&\times
		\frac{\partial^2}{\partial \xi_{\mathrm{B}} \partial \xi_{\mathrm{B}}^*}
		\bigg[
		\exp{\Big(\frac{|\xi_{\mathrm{B}}|^2}{2}\Big)}
		f(\xi_{\mathrm{A}}, \xi_{\mathrm{B}}, \sqrt{\kappa_\mathrm{PC}})
		\bigg]\bigg\rbrace\\
		&-q
		\exp{\Big(\frac{|\xi_{\mathrm{B}}|^2}{2}\Big)}
		\frac{\partial}{\partial \xi_{\mathrm{B}}}
		\bigg\lbrace
		\exp{\Big(-|\xi_{\mathrm{B}}|^2\Big)}\\
		&\times
		\frac{\partial}{\partial \xi_{\mathrm{B}}^*}
		\bigg[
		\exp{\Big(\frac{|\xi_{\mathrm{B}}|^2}{2}\Big)}
		f(\xi_{\mathrm{A}}, \xi_{\mathrm{B}}, \sqrt{\kappa_\mathrm{PC}})
		\bigg]\bigg\rbrace\\
		&-q
		\exp{\Big(\frac{|\xi_{\mathrm{B}}|^2}{2}\Big)}
		\frac{\partial}{\partial \xi_{\mathrm{B}}^*}
		\bigg\lbrace
		\exp{\Big(-|\xi_{\mathrm{B}}|^2\Big)}\\
		&\times
		\frac{\partial}{\partial \xi_{\mathrm{B}}}
		\bigg[
		\exp{\Big(\frac{|\xi_{\mathrm{B}}|^2}{2}\Big)}
		f(\xi_{\mathrm{A}}, \xi_{\mathrm{B}}, \sqrt{\kappa_\mathrm{PC}})
		\bigg]\bigg\rbrace\\
		&+f(\xi_{\mathrm{A}}, \xi_{\mathrm{B}}, \sqrt{\kappa_\mathrm{PC}}),
	\end{aligned}
\end{equation}
where $q=\frac{\kappa_\mathrm{PC}-1}{\kappa_\mathrm{PC}}$.

For the sequential use of the PS and PA operators, the operations PS-PA and PA-PS, we consider the two non-Gaussian operations use the same beam-splitter transmissivity, that is $\kappa_\mathrm{PA}{=}\kappa_\mathrm{PS}$.
For PS-PA, the unnormalized CF is,
\begin{equation}
	\begin{aligned}
		&\bigchi'_{\mathrm{PS-PA}}(\xi_{\mathrm{A}}, \xi_{\mathrm{B}})\\
		&\quad=q^2
		\exp{\Big(\frac{|\xi_{\mathrm{B}}|^2}{2}\Big)}
		\frac{\partial^2}{\partial \xi_{\mathrm{B}} \partial \xi_{\mathrm{B}}^*}
		\bigg\lbrace
		\exp{\Big(-|\xi_{\mathrm{B}}|^2\Big)}\\
		&\quad\times
		\frac{\partial^2}{\partial \xi_{\mathrm{B}} \partial \xi_{\mathrm{B}}^*}
		\bigg[
		\exp{\Big(\frac{|\xi_{\mathrm{B}}|^2}{2}\Big)}
		f(\xi_{\mathrm{A}}, \xi_{\mathrm{B}}, {\kappa_\mathrm{PS}})
		\bigg]\bigg\rbrace.\\
	\end{aligned}
\end{equation}
The unnormalized CF for PA-PS is,
\begin{equation}
	\begin{aligned}
		&\bigchi'_{\mathrm{PA-PS}}(\xi_{\mathrm{A}}, \xi_{\mathrm{B}})\\
		&\quad=(\kappa_\mathrm{PS}-1)^2
		\exp{\Big(-\frac{|\xi_{\mathrm{B}}|^2}{2}\Big)}
		\frac{\partial^2}{\partial \xi_{\mathrm{B}} \partial \xi_{\mathrm{B}}^*}
		\bigg\lbrace
		\exp{\Big(|\xi_{\mathrm{B}}|^2\Big)}\\
		&\quad\times
		\frac{\partial^2}{\partial \xi_{\mathrm{B}} \partial \xi_{\mathrm{B}}^*}
		\bigg[
		\exp{\Big(-\frac{|\xi_{\mathrm{B}}|^2}{2}\Big)}
		f(\xi_{\mathrm{A}}, \xi_{\mathrm{B}}, {\kappa_\mathrm{PS}})
		\bigg]\bigg\rbrace.\\
	\end{aligned}
\end{equation}

\subsubsection{Quantum scissors}
The unnormalized CF of a TMSV state after the QS operation has been applied to mode $\mathrm{B}$ is
\begin{align}
\bigchi'_\mathrm{QS}(\xi_A, \xi_B) =  \Big[ \bigchi_{\ket{0}\bra{0}}(\xi_B) \int \frac{d^2\xi}{\pi} \bigchi_\mathrm{TMSV}(\xi_A, \xi) \bigchi_{\ket{0}\bra{0}} (-\xi)  \nonumber \\
+g_\mathrm{QS} \bigchi_{\ket{0}\bra{1}}(\xi_B) \int \frac{d^2\xi}{\pi} \bigchi_\mathrm{TMSV}(\xi_A, \xi) \bigchi_{\ket{0}\bra{1}} (-\xi) \nonumber \\
+g_\mathrm{QS} \bigchi_{\ket{1}\bra{0}}(\xi_B) \int \frac{d^2\xi}{\pi} \bigchi_\mathrm{TMSV}(\xi_A, \xi) \bigchi_{\ket{1}\bra{0}} (-\xi) \nonumber \\
+g_\mathrm{QS}^2 \bigchi_{\ket{1}\bra{1}}(\xi_B) \int \frac{d^2\xi}{\pi} \bigchi_\mathrm{TMSV}(\xi_A, \xi) \bigchi_{\ket{1}\bra{1}} (-\xi) \Big] \frac{1}{1+g_\mathrm{QS}^2},
\end{align}
where $g_\mathrm{QS}{=}\sqrt{1+\kappa_\mathrm{QS}}/\kappa_\mathrm{QS}$ is the amplification parameter, and $\bigchi_{\ket{m} \bra{n}}$ is the CF of the quantum state $\ket{m}\bra{n}$, obtained as
\begin{align}
\bigchi_{\ket{m} \bra{n}}(\xi)= \bra{m}\hat{D}(\xi)\ket{n} = \sqrt{\frac{n!}{m!}} \xi^{m-n} e^{-|\xi|^2/2} \mathcal{L}^{m-n}_n(|\xi|^2),
\end{align}
where the $\mathcal{L}^{m-n}_n$ is the associated Laguerre polynomial.

We note the above CFs above are normalized by dividing over the norm of the respective CF, viz.,
\begin{equation}\label{eq:nGstate}
	\begin{aligned}
		\bigchi_o(\xi_{\mathrm{A}}, \xi_{\mathrm{B}})&=\frac{1}{\bigchi'_o(0, 0)}
    \bigchi'_o(\xi_{\mathrm{A}}, \xi_{\mathrm{B}})\\
	\end{aligned}
\end{equation}

\subsubsection{Squeezed Bell-like states}
We also consider the non-Gaussian SB states, that are obtained by the application the squeezing operator to a Bell state,
\begin{align}
\ket{\psi}_\mathrm{SB} &=(\cos^2(\delta) + \sin^2(\delta))^{-1/2} \nonumber \\ & \times  \hat{S}_{12}(\varrho) \Big[\cos(\delta) \ket{00} + \sin(\delta)\ket{11} \Big].
\end{align}
The normalized CF of a SB state is \cite{nonGaussianTeleportation}
\begin{align}
&\bigchi_\mathrm{SB}(\xi_\mathrm{A}, \xi_\mathrm{B}) = (\cos^2(\delta) + \sin^2(\delta))^{-1/2}
\nonumber \\
&\exp \Big[-\frac{1}{2}\Big(|\xi_\mathrm{A}'|^2 + |\xi_\mathrm{B}'|^2 \Big) \Big]
  \Big[ \cos^2(\delta)  + 2 \cos(\delta)\sin(\delta)  \nonumber \\
  &\times \Re\{ \xi_\mathrm{A}' \xi_\mathrm{B}'   \} +  \sin^2(\delta) (1 - |\xi_\mathrm{A}'|^2) (1 - |\xi_\mathrm{B}'|^2)  \Big],
 \label{eq:sb}
\end{align}
where $\Re\{ z\}$ is the real part of $z$, and the transformation given by Eq. \ref{eq:Bogo} is used.

\subsubsection{Optimizing the entangled resources}
We emphasize that the key to increasing the effectiveness of quantum teleportation using non-Gaussian states is the optimization of the free parameters in each state such that the fidelity is maximized. It is known that the fidelity is maximized for $\phi{=}\pi$ in the squeezing parameter for all the states considered in this work \cite{DellAnno1}. Moreover, the parameter $g$ in the teleportation must also be optimized. Additionally, the non-Gaussian states possess an additional free parameter which allows further optimization of the teleportation.\footnote{For all the resource states, the squeezing $r$, and the gain $g$ are optimized. For the PA, PS, PC, PA-PS, PS-PA, operations the parameter $\kappa_o$ is also optimized.
For the SB state, the phase $\delta$ is optimized.} Throughout this work, all the parameters in the states used are optimized, and importantly, this is done for each value of the channel transmissivity.

\section{Results}
\begin{figure}
\centering
\includegraphics[width=.46\textwidth]{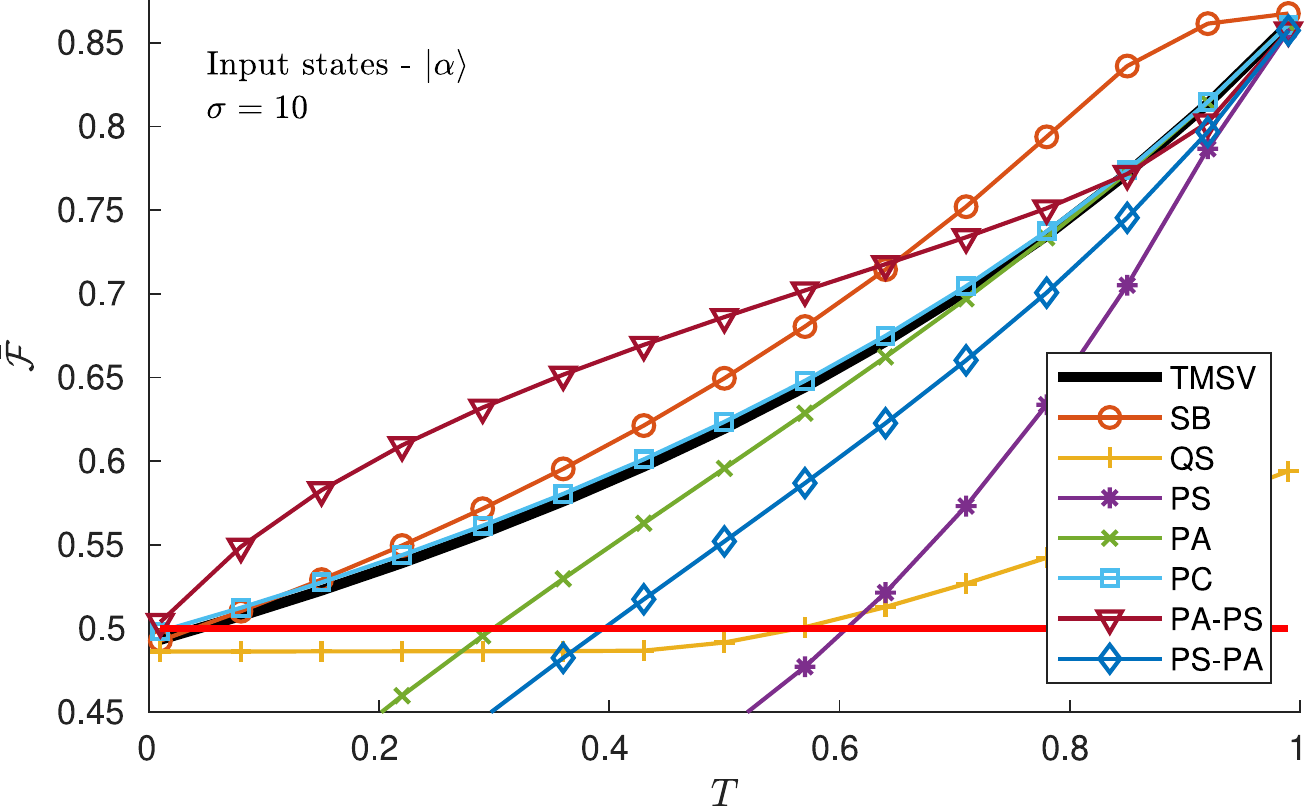}
\includegraphics[width=.46\textwidth]{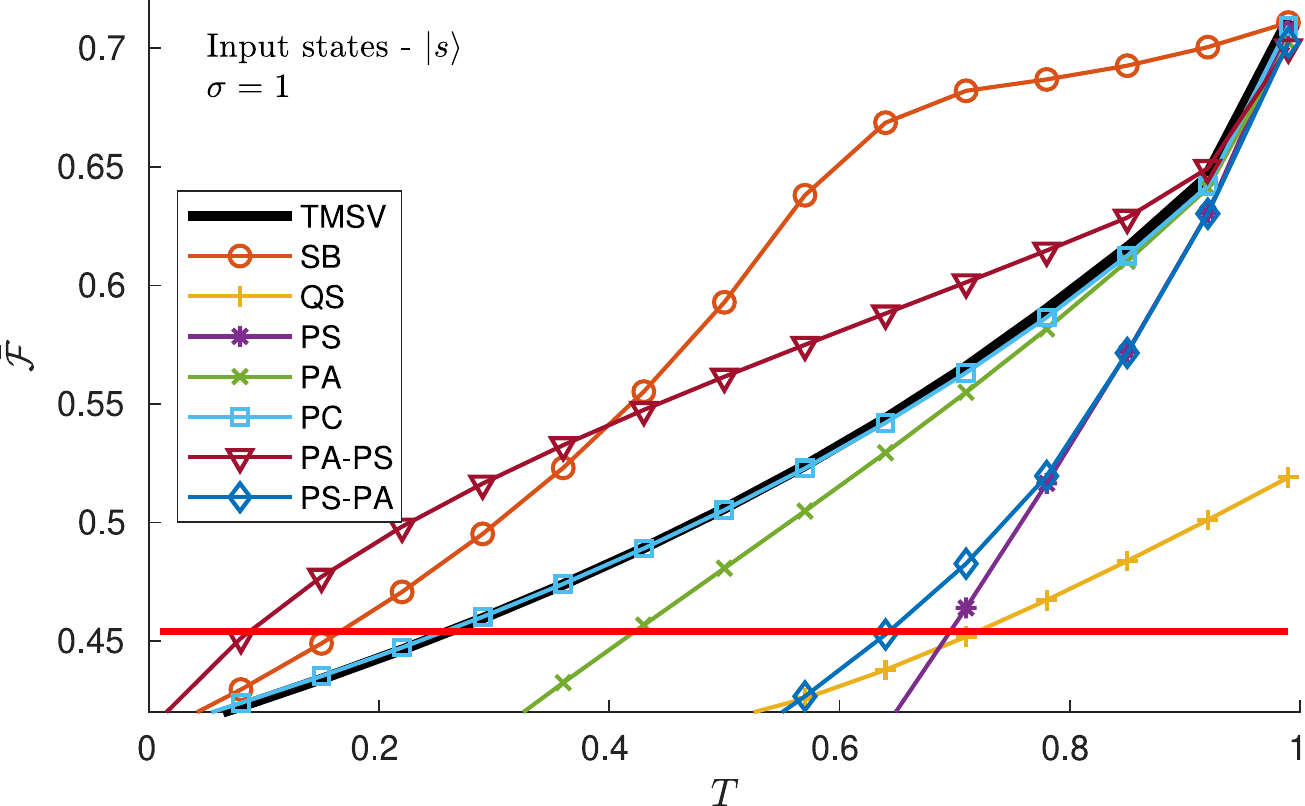}
\caption{Mean fidelity obtained using the non-Gaussian states, and the Gaussian TMSV state for teleported coherent and squeezed states. All the free parameters in the states are optimized to maximize the fidelity. A fixed value of $\epsilon{=}0.05$ is set. The red line represents the classical limit.}
\label{fig:fixed}
\end{figure}

In Fig. \ref{fig:fixed} a comparison of the teleportation fidelities for both input states obtained for all the teleportation resource states is given. The limit on the fidelity that can be achieved by means of local measurements and classical communications, known as the {\it classical limit}, is 1/2 for a single coherent state, and $\sqrt{e^{\overline{\varsigma}}} / (1+ e^{\overline{\varsigma}})$ for squeezed states, with $\overline{\varsigma}$ the mean of $\varsigma$ obtained from the distribution in Eq. \ref{eq:probcoherent} \cite{Owari_2008}. Throughout this work, the efficiency of homodyne measurements is fixed to
$\eta^2{=}1 ~\mathrm{dB}$.
See that several non-Gaussian resources do not present an improvement over the TMSV state, only the SB and PA-PS resource states present higher fidelities.
Counter-intuitively, the PS state under-performs the TMSV state and the PA state, whereas in the case of CV-QKD it has been shown how the use of PS states enhances the key rates \cite{MingjianPAPS}. For the PC state, the fidelity is the same as in the TMSV state. This can be explained by the fact $\kappa_\mathrm{PC}{\rightarrow}1$ during the optimization process, subsequently reducing the state to a TMSV state. For the two states that outperform the TMSV state, the SB and PA-PS states, we observe that for a given $\sigma$ there exists a threshold in the transmissivity below which the PA-PS state yields higher fidelities, and above this threshold, the SB state gives the highest fidelities.

\begin{figure}
\centering
\includegraphics[width=.46\textwidth]{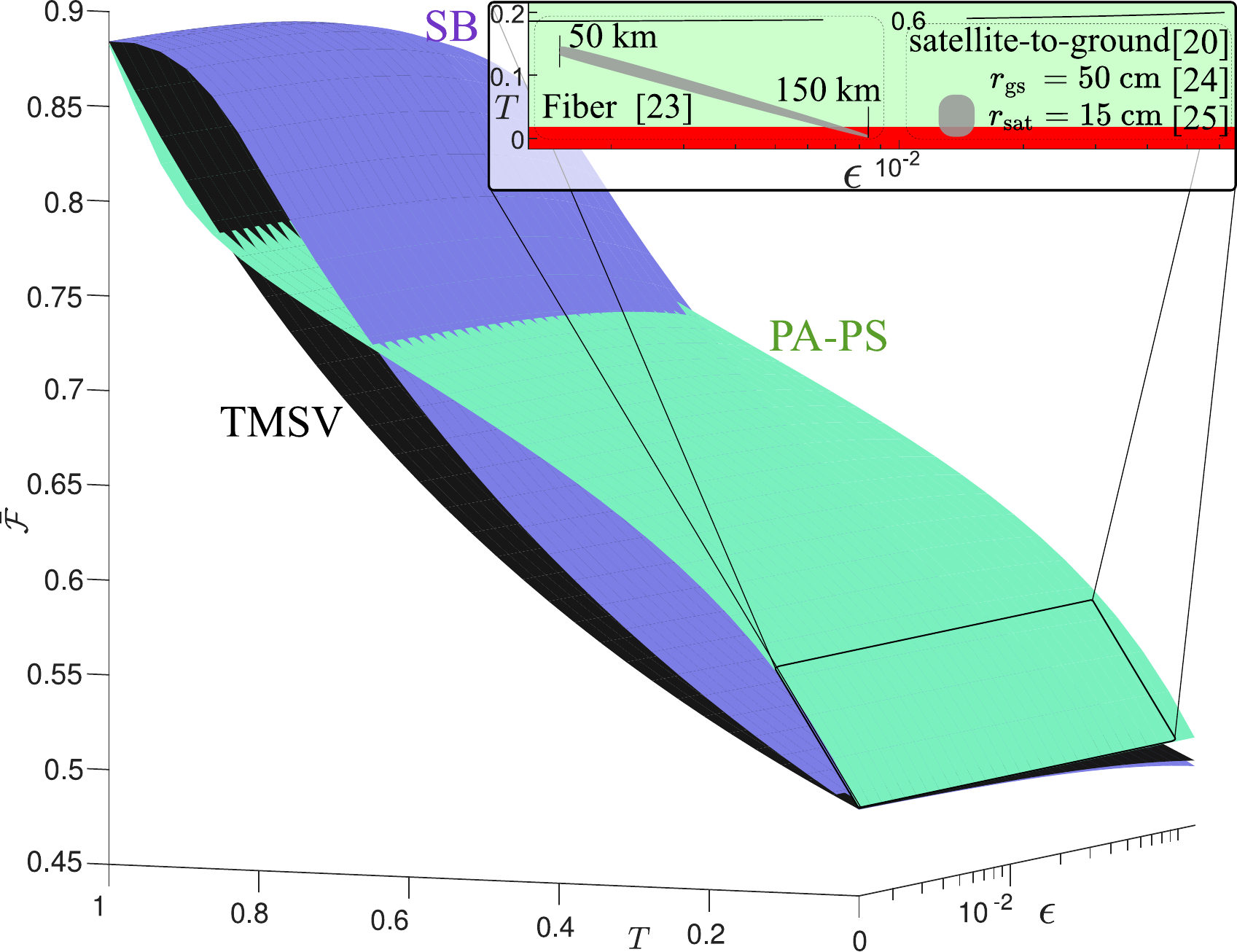}
\caption{Mean fidelity of teleported coherent states as a function of the transmissivity, $T$, and excess noise, $\epsilon$, of the quantum channel. The black, green, and purple curves correspond to the fidelity when the TMSV, PA-PS, and SB states are used as the entangled resource, respectively. The value of $\sigma{=}10$ is used. (inset) The gray area marks the noise parameters associated with realistic quantum channels. The red area corresponds to the classical limit.}
\label{fig:2d}
\end{figure}

\subsection{Performance over realistic quantum channels}
The noise effects of CV quantum channels are specified by  two parameters, ${\epsilon}$ and $T$ (see Eq \ref{eq:CF_loss}).
See in Fig. \ref{fig:2d}, the mean fidelity of teleported coherent states as a function of the two noise parameters. There are two distinct regions in the parameters space where either the SB or the PA-PS states yield the highest fidelities. The regions are mostly determined by the value of the transmissivity, while the excess noise has a considerably lesser impact on the behavior of the fidelity. As before, the SB state only surpasses the PA-PS state for high values of transmissivity, where $T{>}0.72$. We note that in Fig. \ref{fig:2d} we used $\sigma{=}10$, for higher values of $\sigma$ we found no considerable difference with the fidelities presented here.

With the results of Fig. \ref{fig:2d} in mind, we now assess the advantage gained by using non-Gaussian resources on specific real-world quantum channels. We focus on terrestrial state transfer via an optical fiber channel, and space-based state transfer via the satellite-to-ground channel. For the optical fiber channel, current state-of-the-art fiber technology sets the loss at 0.16 dB/km, and an excess noise ranging from 0.0033 for a distance $L{=}50$ km, to 0.0086 for $L{=}150$ km \cite{fiber_loss}. Based on these experimental values we approximate the excess noise as a linear function using the following equation $\epsilon_\mathrm{fiber}{=}aL+ 6{\times}10^{-4}$, with $a{=}5.3 {\times} 10^{-5}\mathrm{km}^{-1}$.

For the satellite-to-ground quantum channel, the properties of the channel are highly dependent on the hardware characteristics of the receiver and transmitter, most notably, on the aperture sizes of the
receiving and transmitting telescopes.
We consider realistic apertures of radii $r_\mathrm{sat}{=}15$ cm and $r_\mathrm{gs}{=}50$~cm, for the satellite (transmitter) and ground-station (receiver), respectively, as in \cite{liao2017satellite}.
The satellite and the ground station have the ability to send and measure quantum optical signals, respectively.
The ground station is positioned at ground level, $h_0{=}0$ km, and
the satellite when directly overhead at an altitude $H{=}500$ km. The total propagation length, between the satellite and the ground station, depends on the zenith angle of the satellite relative to the ground station. For further details see \cite{villasenor2021enhanced}. In the satellite-to-ground channel, the fluctuations of the atmosphere translate to variating transmissivities that have to be sampled from a probability distribution. With the specified aperture sizes, the mean transmissivity of the satellite-to-ground channel ranges from $\overline{T}{\approx}0.06$ at distance $L{=}500$ km, to $\overline{T}{\approx}0.002$ at $L{=}1460$ km. On satellite-based quantum channels, the excess noise not only arises from the hardware involved in the creation and measurement of the signals, but also from the fluctuations the turbulent atmosphere induces on the quantum signals as they are transmitted. The magnitude of the excess noise is dependent on the specific hardware used (i.e. aperture sizes), several theoretical studies set the excess noise in the range of $\epsilon{\in}[0.014 - 0.015]$ \cite{Sebastian,kish2020use}, depending on the total propagation length.

\begin{figure}
\centering
\includegraphics[width=.45\textwidth]{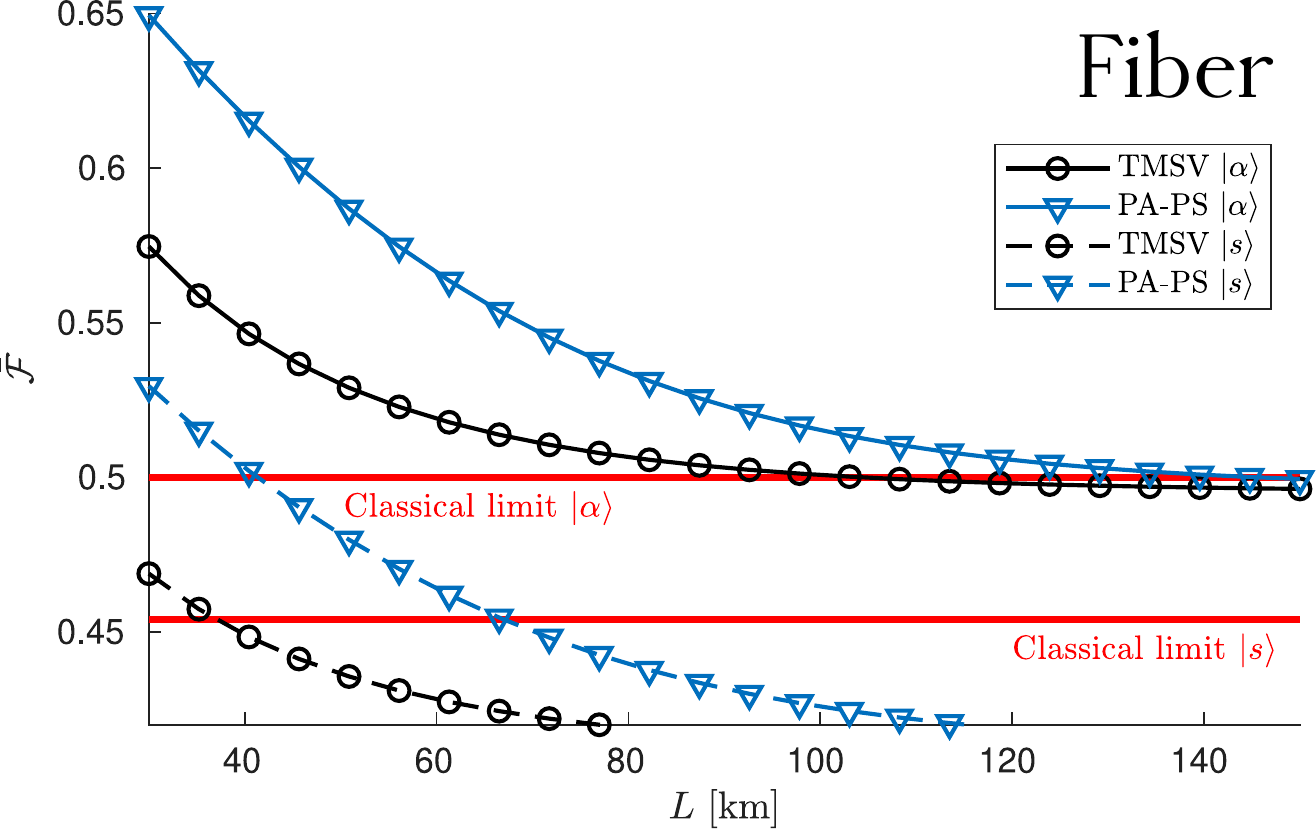}
\includegraphics[width=.45\textwidth]{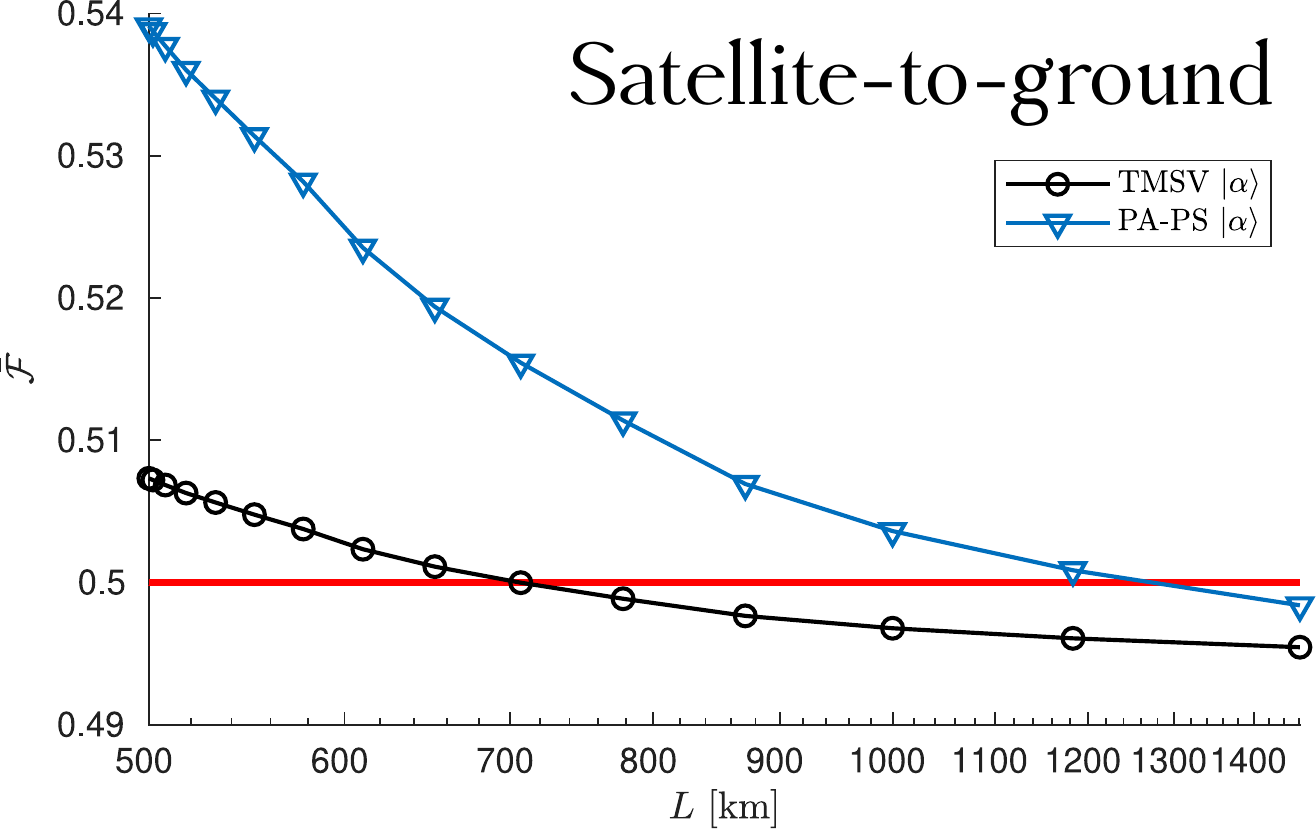}
\caption{Mean fidelity obtained for realistic quantum channels. A value of $\sigma{=}10$ is used in the calculation of the mean fidelity of teleported coherent states, and $\sigma{=}1$ for squeezed states.}
\label{fig:sat}
\end{figure}

In Fig. \ref{fig:sat} we show the advantage obtained by using the PA-PS state on the fiber and satellite-to-ground channels.
In both cases, the use of non-Gaussian resources improved the robustness of the quantum teleportation to the loss of the channel. In the fiber channel, the increase of fidelity is modest, below $15\%$. Nonetheless, the use of the non-Gaussian resource allows the fidelity of teleported coherent states to overcome the classical limit up to 140 km, compared to the limit of 100 km when the TMSV state is used as the resource.
Similarly, when squeezed states are teleported, the maximum distance where the fidelity overcomes the classical limit is increased from 38 km to 65 km by using the non-Gaussian resource.
For teleported coherent states via the satellite-to-ground channel, the increase of the distance is even higher when the non-Gaussian resource is used. Using the TMSV state the fidelity drops below the classical limit for distances larger than 700 km, while with the non-Gaussian resource the fidelity surpasses the classical limit up to ${\approx}1220$ km, an increase of more than $80\%$. In the satellite-to-ground channel, the fidelity of teleported squeezed states was always below the classical limit for the distances considered here.

Finally, we emphasize the importance of parameter optimization in any enhancement of fidelity. We repeated the calculation of the fidelities of teleported coherent states for the realistic channels using the PA-PS state, now with a fixed (non-optimized) value of $r$ and a fixed $g{=}1/\eta$. Under these conditions, regardless of the chosen value of $r$, the distance where the teleportation fidelity surpasses the classical limit never exceeds $92$ km for the fiber channel, and $610$ km for the satellite-to-ground channel.

\section{Conclusions}
We considered CV quantum teleportation of coherent and squeezed states via a wide range of non-Gaussian entangled resource states.
 Of the non-Gaussian resources considered two achieved higher teleportation fidelities relative to the Gaussian TMSV state, namely,  the SB state and the PA-PS state. For quantum channels with high transmissivity (low loss),  the SB state yielded the highest fidelities over any other state considered. Conversely, for channels with low transmissivities,  the PA-PS state surpassed all the other states considered. In many quantum communication implementations such as terrestrial implementations based on optical fiber and space-based implementations, the transmissivity is generally low, making the PA-PS resource the state of choice. In such implementations, the advantage over a TMSV state of using the PA-PS state translates as a significant increase in the distance over which the fidelity of teleportation surpasses the classical limit.

\bibliographystyle{unsrt}


\begingroup
\raggedright

\bibliography{bib}
\endgroup

\end{document}